 \documentclass[final,5p,times,twocolumn]{elsarticle}

\usepackage{lineno,hyperref}
\modulolinenumbers[5]

\journal{Journal of \LaTeX\ Templates}

\usepackage{numcompress}
\usepackage{comment}

\usepackage{epsfig}

\usepackage{amssymb}

\usepackage{float}

\journal{Nuclear Physics A}

\begin{document}

\begin{frontmatter}



\title{Ultra-Compact Ka-band linearizer for the Ultra-Compact X-Ray Free-Electron Laser at UCLA}


\author{B. Spataro$^{a}$, M. Behtouei$^{a}$, L. Faillace$^{a, d}$, A. Variola$^{a}$, V.A. Dolgashev$^{b}$,  J. Rosenzweig$^{c}$, G. Torrisi$^{f}$, and M. Migliorati$^{d, e}$\\\vspace{6pt}}

\address{$^{a}${INFN, Laboratori Nazionali di Frascati, P.O. Box 13, I-00044 Frascati, Italy};\\
 $^{b}${SLAC- National Accelerator Laboratory, Menlo Park, CA 94025, USA};\\
 $^{c}${Department of Physics and Astronomy, University of California, Los Angeles, California 90095};\\
 $^{f}${INFN, Laboratori Nazionali del Sud, Catania, Italy};\\
  $^{d}${Dipartimento di Scienze di Base e Applicate per l'Ingegneria (SBAI), Sapienza University of Rome, Rome, Italy};\\
  $^{e}${INFN/Roma1, Istituto Nazionale di Fisica Nucleare, Piazzale Aldo Moro, 2, 00185, Rome, Italy }
 }

\begin{abstract}
There is a strong demand for accelerating structures able to achieve higher gradients and more compact dimensions for the next generation of linear accelerators for research, industrial and medical applications.

Notably innovative  technologies will permit compact and affordable advanced accelerators as the linear collider and X-ray free-electron lasers (XFELs) with accelerating gradients over twice the value achieved with current technologies. In particular XFEL is able to produce coherent X-ray pulses with peak brightness 10 orders of magnitude greater than preceding approaches,  which has revolutionized numerous fields through imaging of the nanoscopic world at the time and length scale of atom-based systems, that is of femtosecond and Angstrom. There  is a strong interest for combining these two fields, to form a proper tool with the goal of producing a very compact XFEL in order to investigate multi-disciplinary  researches  in chemistry, biology, materials science, medicine and physics.

In the framework of the Ultra -Compact XFEL project (UC-XFEL) under study at the University of California of Los Angeles (UCLA), an ultra high gradient higher harmonic radio-frequency (RF) accelerating structure for the longitudinal space phase linearization is foreseen. To this aim,  a Ka-Band linearizer (34.2 GHz) with an integrated voltage of at least 15 MV working on 6th harmonic with respect to the main Linac frequency (5.712 GHz) is required. We here present the electromagnetic design of a cold ultra compact Ka-band standing wave (SW) linearizer, 8 cm long, working on $\pi$ mode with an ultra high accelerating gradient (beyond 100 MV/m) and minimum surface electric field for minimizing  the probability of RF breakdown. Moreover, we discuss a traveling-wave (TW) option and compare it with the initial SW structure, by means of main RF parameters as well as beam-dynamics considerations. The numerical electromagnetic studies have been performed by using the well known SuperFish, HFSS and CST computing software.

\end{abstract}




\end{frontmatter}


\section{Introduction}

The development of ever more progressed accelerating structures \cite{MM1,MM2} is one of the driving action of the accelerator community. High gradients, efficiency and unwavering quality of the accelerating structures play a fundamental role on linear accelerators. There is a strong demand of these accelerating structures able to achieve higher gradients and more compact dimensions for the next generation of linear accelerators for research in chemistry, biology, materials science, medicine and physics. As an example, a XFELS photoinjector with a high field level of $>$250 MV/m over a factor of 25 increased beam brightness, would have ~1/3 the undulator length, and 3 times the peak power. Such an advance in performance is a paradigm shift in FELs and could allow, for example, new direct electron beam imaging techniques such as ultra-relativistic electron diffraction and microscopy. 

\begin{figure*}[t]

\begin{center}
\fbox{ \includegraphics[width=0.9 \textwidth ]{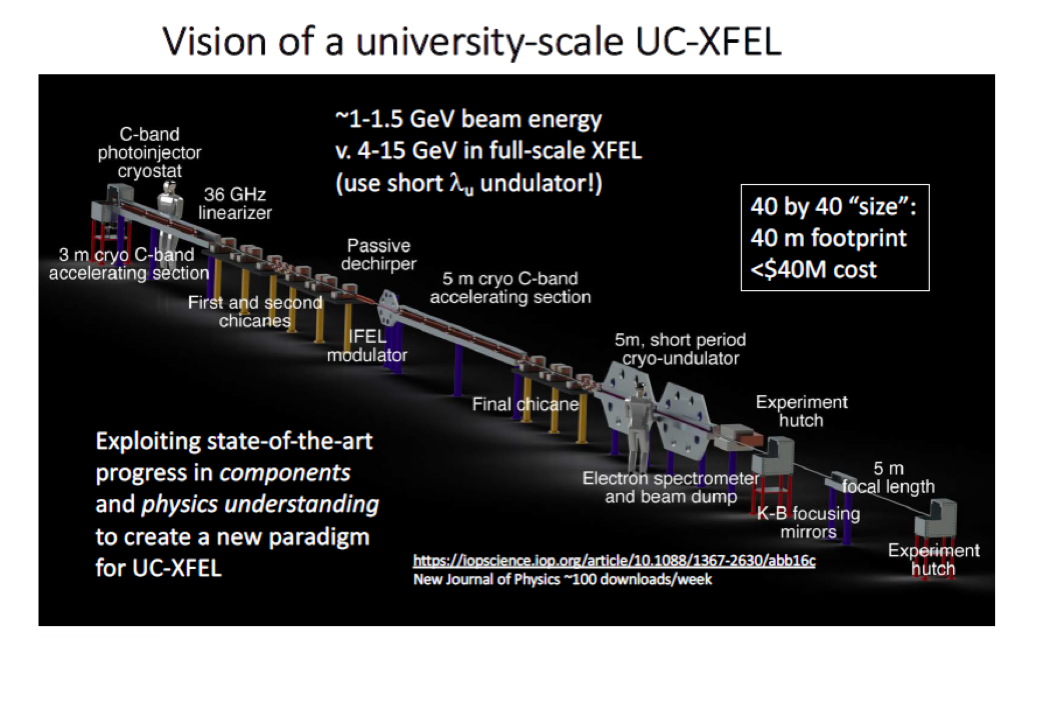}}
 
 \end{center}

\caption { Schematic layout of the Ultra-Compact X-Ray Free-Electron Laser of UCLA.   }\label{fig:layout}
    
     \end{figure*}

The UCLA Particle Beam Physics Laboratory program (PBPL) has established a nearly 30-year track record, as a leading university-based group in beam and accelerator research in both experiment and theory. Its multi-disciplinary approach permits study over a wide spectrum of frontier topics, ranging from very high brightness electron sources using sophisticated microwave structure (RF photoinjectors) \cite{ref3,ref4,MM3,MM4}, to high gradient wakefield acceleration in plasmas and dielectrics, and extending to fundamental ultra-fast beam physics and diagnostics, and beyond, to advanced light sources such as free-electron lasers and Compton scattering sources necessary for wide variety of applications \cite{MM6,ref5,ref6}. Recent progress in very high gradient radio-frequency acceleration techniques has been carried out in recent years by a UCLA-SLAC collaboration \cite{ref1}, with notable contributions also arising from INFN-LNF and LANL~\cite{ref_New1,ref_New2,ref_New3,ref_New4, ref_New5}. 

With this work we considered a new technique, that of cryogenic cooling of RF cavities, successfully at the frontier of achievable gradients using microwave accelerators, with maintenance of surface electric fields up to 500 MV/m \cite{ref2}. On the linac application side, SLAC has recently produced a design study of a collider based on cryogenic high gradient RF acceleration. This study points out that the limits of performance profoundly changed in cryogenic operation, and C-band, with its attendant lower wakefields, may be the optimal RF frequency choice for a  $>$100~MeV/m linear accelerator. Such accelerators may be employed for compact FELs, Compton sources, wakefield accelerators, and many other applications. Indeed, due to the higher frequency and lower thermal load, C-band is also favored for many photoinjector applications \cite{ref3,ref4,MM7}, particularly in FEL due to the desire for very low emittance.

This project would represent the first use of such an accelerator structure operating at cryogenic temperatures. It is also the initial development of a cryogenic, ultra-high brightness RF photoinjector. Together, these structures permit testing of the system with a high level of beam loading, by exciting a bunch train from the photocathode. The low loss of the system implies a high RF-to-beam power efficiency, in excess of 60$\%$ \cite{ref5}. The UC-XFEL machine to be constructed at UCLA \cite{ref6}, includes  a cryogenic short C band linac able of providing a beam energy of 300 MeV with an emittance 0.2 mm-mrad. A conceptual layout is reported in Figure~\ref{fig:layout} of \cite{ref6}, and it shows the necessity of a Ka-Band RF structure for correcting longitudinal beam phase-space non-linearity as arises in the C band linac in an ultra compact X-ray FEL is required. There are no dedicated dampers of the parasitic higher order modes for the linearizer structure because the beam dynamics is not influenced by the long-range wake-fields due to the foreseeing  of the single bunch operation. 

Several laboratories worldwide, such as the Laboratori Nazionali di Frascati (INFN-LNF) and UCLA, are involved in the modeling, development and tests of RF structures devoted to particles acceleration with higher gradient electric field through metal device, minimizing the breakdown and the dark current. In particular, new manufacturing techniques for hard-copper structures are being investigated in order to determine the maximum sustainable gradients well above 100 MV/m and extremely low probability of RF breakdown. The preferences of utilizing high frequency accelerating structures are well known: smaller size, higher shunt impedance, higher breakdown threshold level and short filling time. In refs. \cite{ref7,ref8,ref9,ref10,ref100,ref13,ref14,ref16}, it is shown that there are reasonable candidates for microwave tube sources, which, together with RF pulse compressor (SLED), are capable of supplying the required RF power. The technologies in the Ka-Band accelerating structures, high power sources and modulators have also been developed by getting very promising results in order to reach a RF power output of (40-50) MW by using the SLED system.

RF stability during operation and tuning tolerances are important points for the RF structure design in the high frequency range. The complexity of machining, tight mechanical tolerances and alignments are therefore important aspects which have to be taken into account in the design activity. In order to obtain a longitudinal phase space linearization we have designed a compact sixth harmonic standing wave (SW) accelerating structure operating at a frequency of F= 34.272 GHz working on the $\pi$ mode at about 100 MV/m accelerating gradient. This report discusses the fundamental RF parameters dependence of the structure as function of the iris's aperture and the cavity's geometry in order to get the optimized RF design. The figures of merit, like the longitudinal shunt impedance, quality factor, coupling coefficient, transit time factor and so on, are calculated with the well known Superfish, HFSS and CST numerical computing software \cite{refSuperfish,refAnsys,refCst}. Moreover, in order to estimate the breakdown thresholds, investigations on the modified Poynting vector and  pulse heating are discussed. A comparison with a linearizer traveling wave structure operating on the 2$\pi$/3 is discussed too.  Finally operational limit due to RF breakdowns at higher power operation \cite{ref14,ref16} is presented.

\section{Motivation}

One of the most important requirements, for any accelerating structure design, is related to the stability of the particles in the beam.
Like in any RF accelerating structure, the build-up of the wakefields and corresponding wake potentials generated from the beam-cavity interaction and acting back on the bunch itself, can be destructive, leading to the beam break-up (BBU). Our concern regards here the transverse wakefield, since the beam has very high energy ($\ge$ 300~MeV) and it is ultra-short (350~fs rms). In our project, operation is foreseen in single-bunch regime, therefore the only concern is due to the effect of the so-called single-bunch BBU. We will discuss the effect of multi-bunch BBU in a forthcoming paper.

Here, we give an analytical estimation of the wake potential inside the Ka-Band cavity by using the diffraction theory~\cite{Bane}. The transverse wakefield of a pillbox cavity of radius $b$, gap $g$ and iris aperture radius $a$ under the hypothesis that 
\begin{equation}
\label{eq:1}
    \frac{2g\sigma}{(b-a)^2}\ll1
\end{equation}
with $\sigma$ the rms bunch length, can be written as
\begin{equation}
    w_{\perp}(z)= \frac{2^{3/2} Z_0 c}{\pi^2 a^3}\sqrt{gz}
\end{equation}
where $Z_0$ is the vacuum impedance and $c$ the speed of light. In our case, the ratio of Eq.~(\ref{eq:1}) is about 0.13 in the worst case of standing wave structure (having the largest gap of 4.374 mm) with an iris radius of 1.33 mm and the cavity radius of 3.8776 mm.

This wakefield can be convoluted with a Gaussian or rectangular bunch to obtain the corresponding wake potentials. For example, the wake potential of a rectangular bunch of length $l_0$ can be written as
\begin{equation}
    W_{\perp}(z)= \frac{2^{5/2} Z_0 c}{3\pi^2 a^3}\sqrt{gl_0} \left(1+\frac{z-l_0/2}{l_0}\right)^{3/2}
\end{equation}

\begin{figure}[ht]
\begin{center}
\label{fig:wakepot}
\includegraphics[scale=0.5]{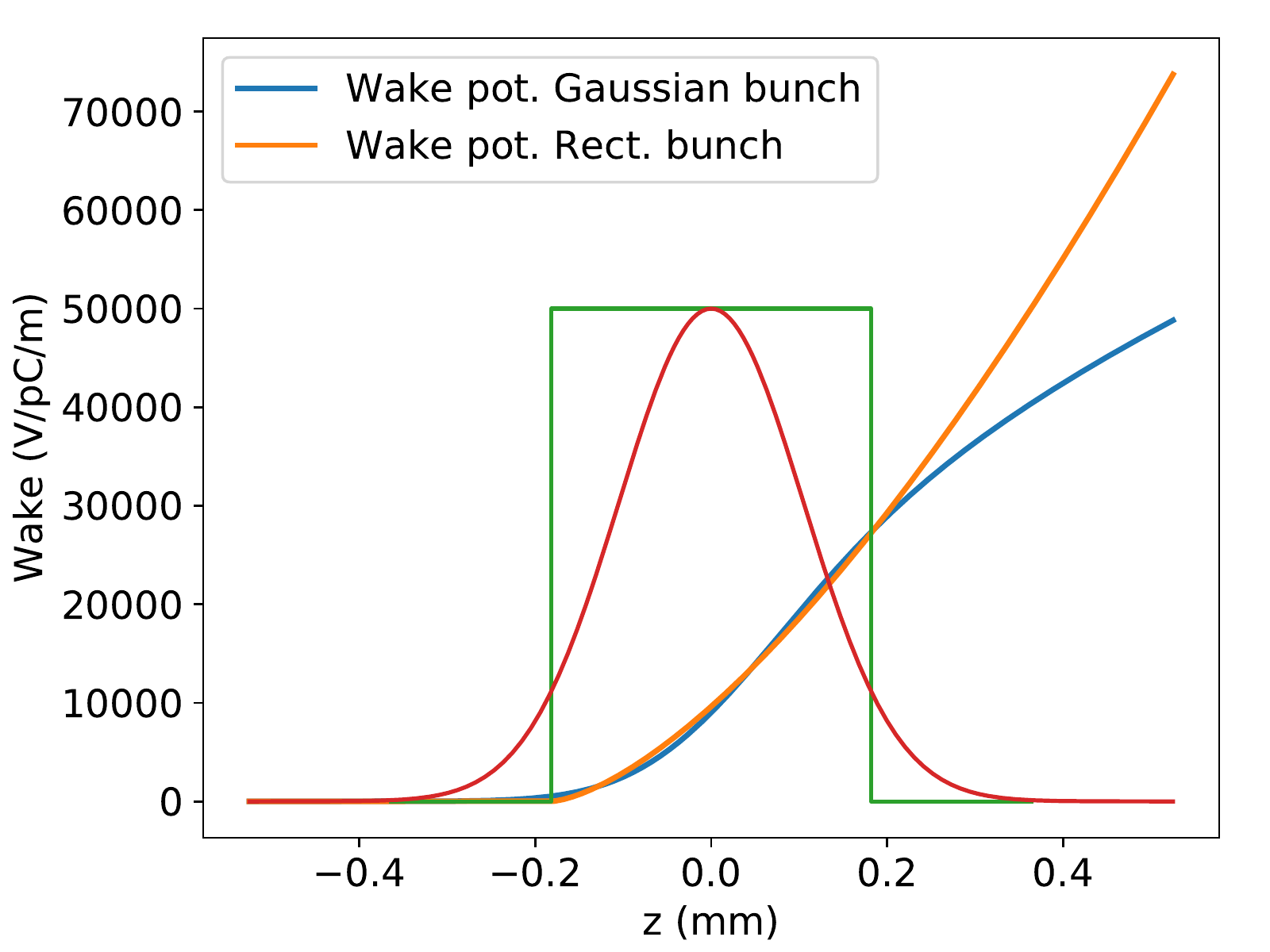}
\caption {Transverse wake potential of a Gaussian and rectangular bunch shape for the wakefield obtained with the diffraction theory.}
\end{center}
\end{figure}

The wake potentials for the two bunch shapes are shown in Fig.~\ref{fig:wakepot}. For the rectangular case we have used a total bunch length equal to $l_0=\sqrt{12}\sigma$.

We can see from the figure that the wake is similar in both cases and it can be considered linear inside the bunch. As a consequence, for the analytical approach here described, we use the linear approximation for the wake potential and a rectangular bunch distribution.

To evaluate the strength of the BBU, we consider two extreme cases. Under the further assumptions of mono-energetic beam and constant beta function $\beta$ inside the structure of total length $L$, we can define the parameter
\begin{equation}
    \eta=\frac{\beta Q W_{\perp}(l_0/2) L}{g E/e}
\end{equation}
where $Q$ is the bunch charge, $E$ is the bunch energy, $e$ is the electron charge, and the wake potential is evaluated at the bunch tail. If $\eta \gg 1$, that is in case of very strong BBU, we can use the following asymptotic expression~\cite{MM8} to evaluate the final transverse displacement of the tail ($x_f$) with respect to the initial one ($x_i$)
\begin{equation}
\label{eq:bbu}
    \frac{x_f}{x_i}=\sqrt{\frac{1}{6\pi}}\eta(L)^{-1/6}e^{\frac{3\sqrt{3}}{4}\eta(L)^{1/3}}
\end{equation}

On the other hand, if $\eta \ll 1$, that is in the weak BBU regime, we can use a perturbation method to obtain the first order solution of the BBU equation, that, in case of constant energy, gives simply
\begin{equation}
    \frac{x_f}{x_i}=\frac{\eta}{2}
\end{equation}

\begin{figure}[ht]
\begin{center}
\label{fig:bbu}
\includegraphics[scale=0.5]{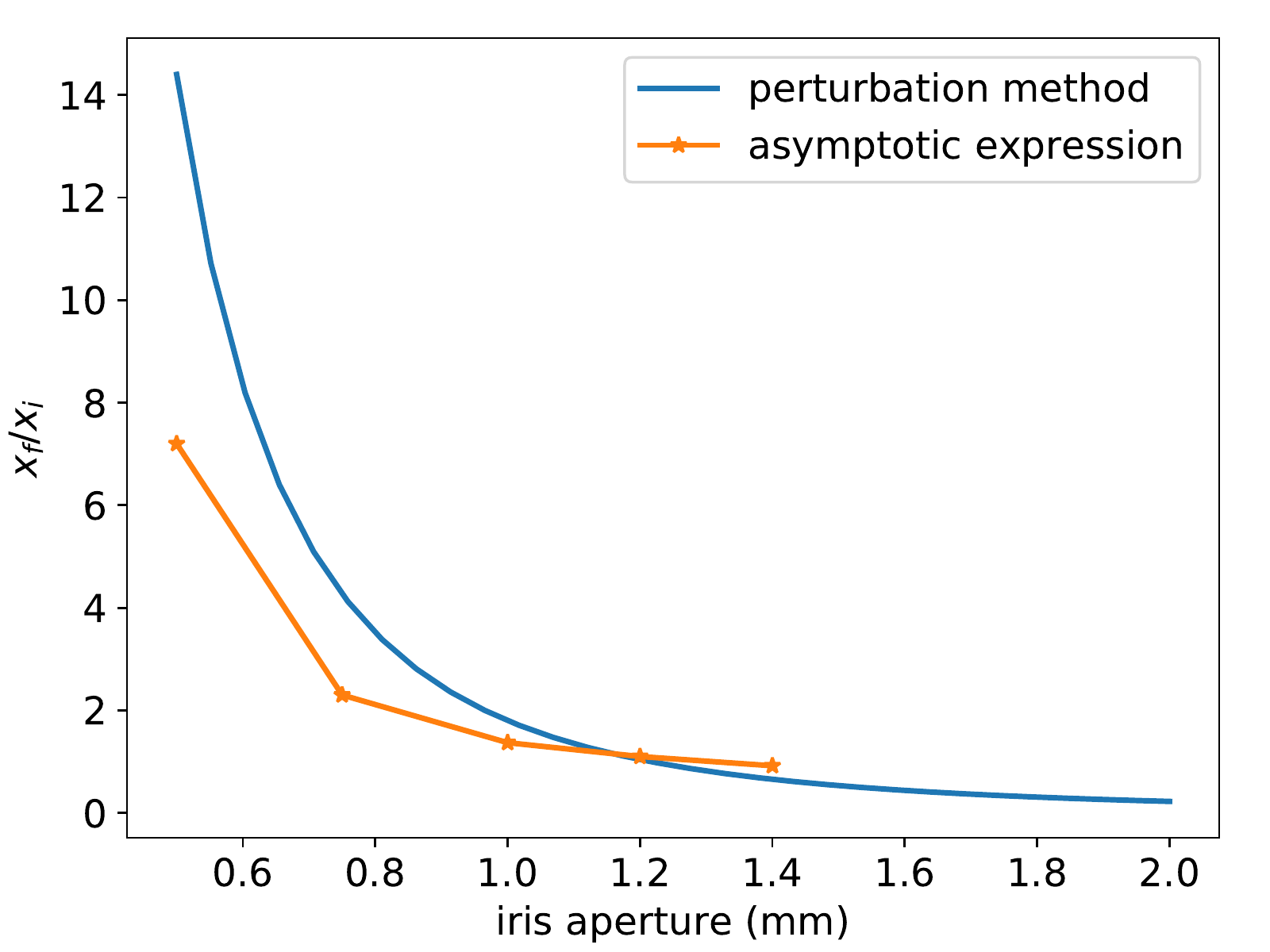}
\caption { bbu}
\end{center}
\end{figure}

In Fig.~\ref{fig:bbu} we show the final transverse displacement of the tail with respect to the initial one for both the cases of strong and weak BBU as a function of the iris aperture in order to find an optimal working point for the case of the standing wave structure of total length of 16 cm.

We observe that in the perturbation method case, since $\eta$ is linear with the transverse wake potential and with the machine length, we can easily deduce the BBU dependence by the different parameters. In particular the BBU increases inversely to the third power of the iris radius and to the square root of the cavity gap. Therefore, for a travelling wave structure, we expect an increase of the BBU  by a factor of about 20 \% with respect to the standing wave one. More complicated is the behaviour in case of strong BBU. However, around an iris aperture of about 1 mm, we obtain similar results. In particular, for a total cavity length of $L$ = 16~cm, we observe a ratio of about 2 which means that for an initial offset of the injection bunch of 100 um, the tail is displaced at the exit by about 200 um, which is still a factor of 4 lower than the iris radius.

As for the longitudinal plane, in case of a rectangular shape, the wake potential can be written as:
\begin{equation}
    w_{||}(z)= \frac{2^{1/2} Z_0 c}{\pi^2 a}\sqrt{\frac{g}{l}}\sqrt{1-\frac{z-l/2}{l}}
\end{equation}

In both SW and TW cases, one obtains a loss factor per unit length equal to nearly 800 keV/m, which is a negligible value with respect to the beam operation energy.

\begin{figure}[ht]
\begin{center}
\label{fig:Pwang_TW_34GHz}
\includegraphics[scale=0.21]{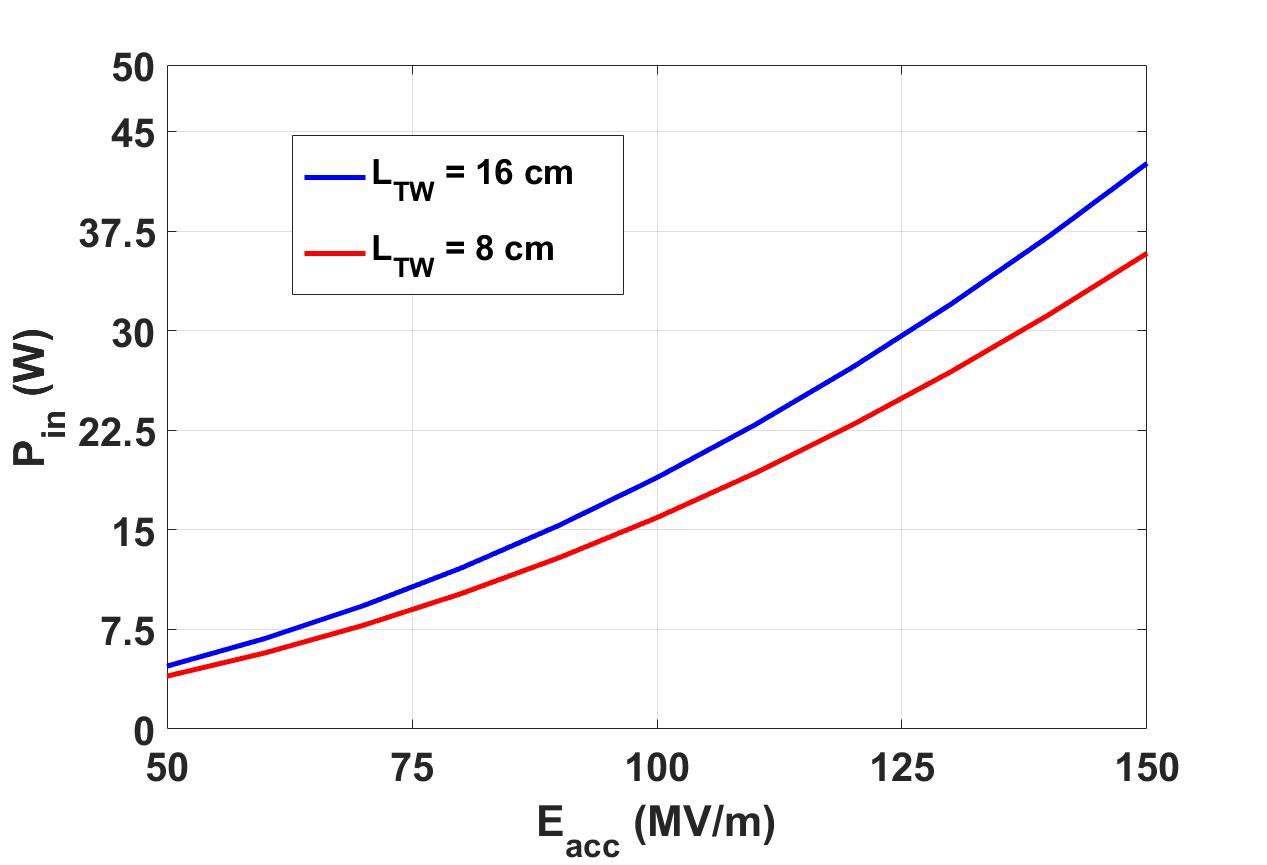}
\caption {Required input RF power vs. accelerating gradient for the TW structure in the case of two lengths, 8 cm and 16 cm. $E_{acc}$ corresponds to an integrated voltage of 20 MV.}
\end{center}
\end{figure}

By choosing a length of 16 cm and an iris radius of 1 mm, we then provide the evaluation for the best performing type of accelerating scheme, for both standing-wave (SW) or traveling-wave (TW) cases. As published in \cite{ref26}, we demonstrated that the optimal maximum length for the SW structure is around 8 cm, in order to avoid resonant mode overlapping. In SW operation, our proposal is then to utilize two SW structures, each 8 cm long, fed by a 3dB hybrid coupler. This setup is also beneficial for protecting the high-power source and avoiding a high-power circulator, which is not trivial to fabricate in Ka-Band. From simulations, it turns out that the required input RF power for an 8 cm long SW cavity is around 8 MW. On the other hand, for the TW case, the required input RF power is shown in Figure~\ref{fig:Pwang_TW_34GHz} for an 8 cm long structure as well as 16 cm, that is possible only in TW operation. The group velocity is assumed to be an average value in the estimation of the required power for both constant-gradient and constant-impedance TW structures. In any case, the needed RF power for the same 8 cm long structure is a factor of 3 less for the SW option.

Eventually, we give the final comparison of the main RF parameters between a SW and TW structure with a length L = 8 cm in Table~\ref{Tab:Comparison}. The comparison obviously shows that it is reasonable to operate a SW structure, which we have then chosen as our optimal option, by means of an available 3-4 MW gyrotron with a pulse compressor (SLED).

\begin{table}[h]

\caption{Comparison between SW and TW.}\label{Tab:Comparison}

\begin{center}

\small{ \begin{tabular}{|| c| c| c||}
\hline
Main RF parameters & SW & TW \\ 
 \hline\hline
Frequency [GHz]& 34.272& 34.272\\ 
\hline 
Accelerating Gradient [MV/m]& 125&125\\ 
\hline 
Input RF power [MW]& 8 & 23.9 \\ 
\textbf{Input RF power @77K [MW]}& \textbf{3.63} & \textbf{10.8} \\ 
\hline
Attenuation [$m^{-1}$] & - & 2.51\\ 
\hline
Shunt impedance [M$\Omega$/m] & 154 & 158\\ 
\hline
Unloaded quality factor, Q & 5728 & 4110\\ 
\hline
Group velocity [$\%$] & -  & 3.65\\
\hline
Coupling coefficient, K$\%$&0.83 & - \\ 
\hline
Structure length [cm]& 8&8\\
\hline
Build-up [ns]& 12.5 & - \\
\hline
Filling Time, $T_f$ [ns]& - & 7.3 \\
\hline
RF Pulse length flat top [ns]& 50&50\\
\hline
Repetition rate [Hz]& 100&100\\
\hline
Average RF power/m [kW/m]& 0.52 & 0.52 \\
Average RF power/m @ 77K [kW/m]& 0.23 & 0.23\\
\hline
\end{tabular}}
\end{center}
\end{table}

\section{Vacuum RF breakdown}

It is well-known that the vacuum RF breakdown, pulsed surface heating and field emission are among the major issues limiting the highest achievable accelerating gradient inside an RF cavity.

In particular, the RF breakdown instantaneously affects the RF power fed into the accelerating structure, thus lowering its performance~\cite{ref_New1}. Being probabilistic, many experiments have been carried out in order to study and then measure the breakdown-rate probability of such phenomenon. As of today, most of these studies have been performed at X-Band frequencies~\cite{ref_New4,ref_New5,ref_New8}, showing gradients well beyond 100~MV/m. This is the demonstration of the accelerating gradient scaling with frequency.

Material technology and handling represents another crucial factor for the determination of the ultimate performance of an accelerating structure. In order to study the contribution of this factor to the RF beakdown-rate statistics, various technological techniques have been investigated and used here as reference ~\cite{ref_New9,ref_New10,ref_New11,ref_New12, ref_New15}. From these experiments, we have learned that hard-copper alloys, at warm temperature, show the best performance in terms of accelerating gradients. In particular the best performance was obtained with CuAg that reached an accelerating gradient of 200~MV/m at $10^{-3} $/pulse/m breakdown probability using a shaped pulse with a 150 ns flat part. On the other hand, under cryogenics operation (cold temperature below 77K), accelerating gradients up to 250 MV/m with surface fields of 500 MV/m can be reached~\cite{ref1,ref2}.

Therefore, the combination of high RF frequency and low temperature operation will bring the proposed Ka-Band linearizing cavity to the achievement of unprecedented values, both in terms of accelerating gradient and vacuum breakdown rate probability. In the next section, we discuss the details at cryogienic operation of the Ka-Band RF linearizer.

\section{Cryogenic Operation of a Ka-band RF Linearizer}\label{Cryogenic Operation}
 
 The use of cryogenic structures to both diminish the RF dissipation and to mitigate breakdown is by now well established, particularly through testing of X-band and S-band devices. In these experiments, the scaling of RF dissipation according to the theory of the anomalous skin effect (ASE) \cite{ref17} has been verified, and surface fields over 500 MV/m have been achieved before breakdown is observed \cite{ref1}.  The advantage in dissipation effects diminishes somewhat at high frequency, but is still notable up to Ka-band, the operating frequency that has been proposed for a high gradient, compact linearizer. This component is critically important for applications such as the MaRIE XFEL \cite{ref18}, the CompactLight FEL,  the Ultra-Compact XFEL at UCLA \cite{ref6}. Paired with the CompactLight sponsored initiative to develop a 15 MW-class klystron at 36 GHz, a compact, high gradient cryogenic linearizer in this frequency range now seems within reach~\cite{ref26}. 
 
Here we review the scaling laws that allow approximate prediction of the performance of such a linearizer, based on  a derated 5 MW input. To orient the expected performance, we note that the shunt impedance calculated for an optimized 36 GHz structure at room temperature is 158 M$\Omega$/m. We can scale the expected behavior of this shunt impedance from detailed calculations of ASE enhancement at low temperature in S-band by a factor of 5. To extend this to Ka-band, we note that the ohmic model scaling of surface resistivity is $R_{s,\Omega}\propto \omega^{1/2}$, while for ASE, the scaling in the low temperature limit is $R_{s,ASE}\propto \omega^{2/3}$. This means that the expected enhancement  in the quality factor has a scaling $Q_{enh}\propto \omega^{-1/6}$, and for low temperature (below 40 K), one may expect in Ka-band $Q_{enh}\simeq 3.3$. For less ambitious cooling designs, operating with liquid nitrogen at 77 K, we may foresee an enhancement of 2.2. 

To give an idea of what is possible with this approach, we assume, as stated above, a 5 MW matched input into a 10 cm long structure, operating at 77 K, with estimated shunt impedance of 349 $M\Omega/m$. In this case, the accelerating field is 130 MV/m, which is well below the breakdown limit of 250 MV/m. The corresponding surface field of 260 MV/m is also below the threshold of dark-current emission of $\sim$ 300 MV/m that is strong enough to beam load the structure \cite{ref2}. Further, at this frequency, the normalized vector potential is a factor of three below that needed to capture and accelerate dark current, further mitigating potential issues with spurious field emission effects. 

\begin{figure*}[t]
 \begin{center}
\fbox{\includegraphics[scale=0.2]{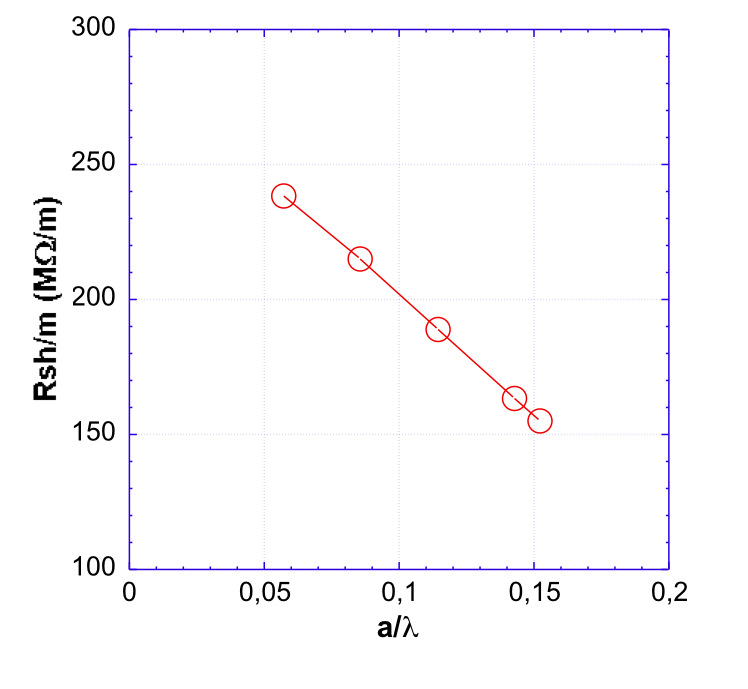}(a)}
\fbox{\includegraphics[scale=0.196]{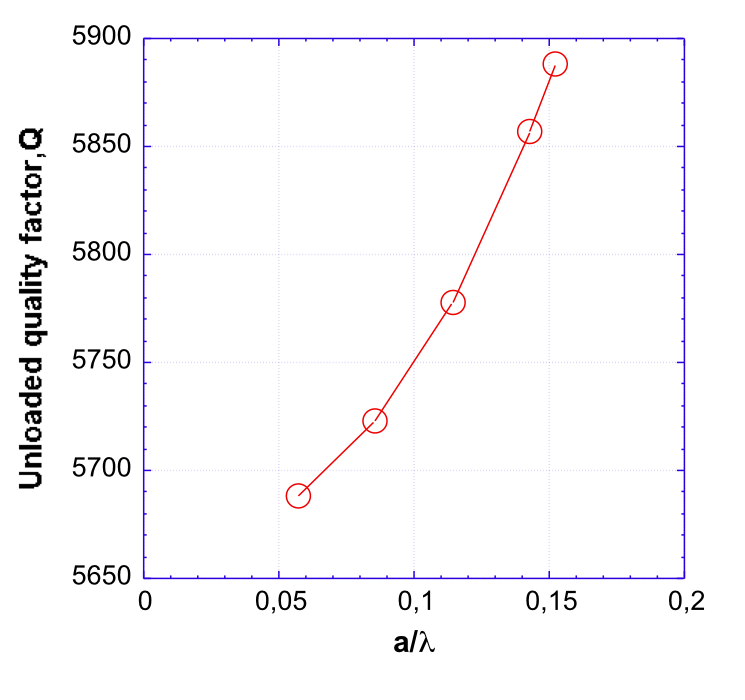}(b)}
\fbox{\includegraphics[scale=0.2]{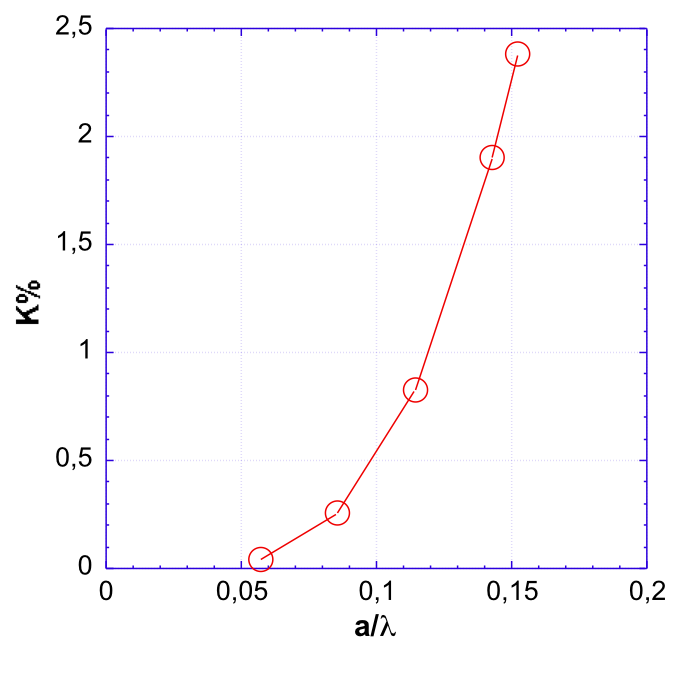}(c)}
\caption { a) Shunt impedance b) unloaded quality factor and  c) coupling coefficient K ($\%$) as a function of the a/$\lambda$ ratio, where a is aperture radius.}\label{fig:Shunt}
\end{center}
\end{figure*}

\section{Cavity choice criteria}
Exhaustive discussions on the design cavity criteria in order to choose  the cavity RF parameters have already described elsewhere \cite{ref26,ref29,ref30} in case of Ka-band structure to be used for the Compact Light XLS project.  

The ultra compact Ka-band linearizer has to fulfill many requirements achievable  as a compromise among the available power source, maximum effective accelerating gradient, simple geometry (for easy fabrication with reasonable tolerances within 2 $\mu$m for getting small sensitivity to construction errors), narrow spread of electron energy, beam dynamics effects in terms of beam loading, beam break-up etc. To minimize the input power required for a given accelerating gradient, the geometry should be designed with the aim of maximizing the effective shunt impedance per unit length. This one occurs with an about  (cavity gap)/(cavity radius) ratio of 1 as it is also shown by using an analytical check of a pill-box cavity. On the other hand, because of the interaction between the beam and the surroundings, the accelerating section performances are limited by effects such as the beam loading, instabilities, beam break-up etc.~\cite{MM8,MM9}. Since we work with a low charge and single bunch, we expect no specific problems on the beam dynamics quality. The third harmonic frequency of the main Linac implies small physical dimensions and thereby the dissipated power constitutes one of the main constraints. A reasonable upper limit on the average power dissipation has been estimated to be the in the  (4-5) kW/m  range \cite{ref22}. To meet the full requirements by keeping a flexibility margin, a structure with simple geometry and of reliable construction with satisfactory mechanical tolerances has to be chosen. 
In this paper, the main concern is to design an accelerating structure operating on the $\pi$ mode with the requirements referring to the Table 1.

 \begin{table}[h]

\caption{Parameters list for the cavity design}

\begin{center}

\small{ \begin{tabular}{|| c| c||}
\hline
Design parameters&  \\ 
 \hline\hline
Frequency [GHz]&  34.272\\ 
\hline 
Effective Accelerating Electric Field [MV/m]  & 125\\ 
\hline
Axial length[cm]&8 \\ 
\hline
Iris Aperture radius (a/$\lambda$)*& 0.114 \\ 
\hline
Iris thickness (h/$\lambda$)& 0.076 \\ 
\hline
 Ratio of phase to light velocity $(\nu_\phi / c)$ & 1 \\ 
 \hline
Pulse charge [pC]& 75\\ 
\hline
Rms bunch length ($\sigma_\tau (fs)$)& 350\\ 
\hline
Pulse repetition rate frequency [Hz]& 100\\
\hline
\end{tabular}}
\end{center}
* a: iris radius of the structure; $\lambda$= free space wavelength
\end{table}

It should be emphasized that in normal conducting operation a 125 MV/m effective accelerating electric field can be  considered as upper limit. Basically only experimental tests can confirm reliability and stable operation to this level of field. For sake of completeness, high power tests on a 100 GHz $\pi$ mode structure, a 10 ns long pulses and a/$\lambda$ = 0.286 mm,  at room temperature, provided an unprecedented high gradient up to  230 MV/m with a peak surface of more than 520 MV/m \cite{ref222}. Technological advancements in terms of precision and assembly of the accelerating structure will allow to reach the design value of 300 MV/m accelerating gradient for 600 kW of dissipated power. As a result, it is reasonable to predict to work in the (100-150) MV/m range in the Ka-band regime. It is also reasonable  that by working at a cryogenic temperature of 77 K degree, the field can reach a level up to 150~MV/m.

\section{Accelerating structure design}
 
The Ka-band structure linearizing the longitudinal phase space has to fulfill many requirements dominated mainly by the lack of high  power klystron amplifier at 36 GHz to be adopted for feeding the linearizer.  The key requirement of the ultra-compact linearizer  Ka-band is to  achieve an ultra high effective accelerating gradient using an available amount of power source with negligible wakefield effects on the beam dynamics quality. Therefore special care should be taken  in order to fulfill the structure aperture for achieving  a high shunt impedance and use the available power source.  Currently, high gradient  structure in the Ka-band regime can be operated with the gyro-klystrons devices which are able to provide an output power up to 3 MW, and with the SLED system which could provide 12 MW by assuming a compression factor of 4. It is well known that the performances of high gradient structures are limited by breakdown due to the pulse heating and field emission effects. 

On the other hand, the RF properties depend strongly on the a/$\lambda$ ratio where a is the iris aperture and $\lambda$ the free space wavelength.  A detailed analysis on  the SW structure operating in the Ka-band regime for the Compact Light XLS project  has been discussed in a recent paper\cite{ref26}. Much care was used in choosing the cavity shape in order to reduce or avoid breakdown effects at a high accelerating gradient operation. As a result, an extensive electromagnetic analysis of the main RF parameters as function of the cavity geometry  has been performed for getting a  RF design with a minimum surface electric field. By assuming an iris thickness h = 0.667 mm \cite{ref22, ref29,ref30}, working on $\pi$ mode, we have  quantified the influence of the iris radius and of its geometry on the main RF parameters. As a result, the RF structure optimization has been reached with a iris radius in the (0.11- 0.14) a/$\lambda$ range in case of elliptical geometry iris,  with a 5/7 semi-axes ratio and with a/$\lambda$ = 0.114, very similar to 11.424 GHz structure and 110 GHz one  \cite{ref222,ref31,ref32,ref34,ref35} for  a direct comparison. We decided to extend a SW structure design also to case of cryogenic operation a 77 K  operating on $\pi$ mode in order to get a satisfactory longitudinal shunt impedance and an acceptable iris aperture for practical beam dynamics considerations.
 
\begin{figure}[h]
\begin{center}
\fbox{\includegraphics[scale=0.25]{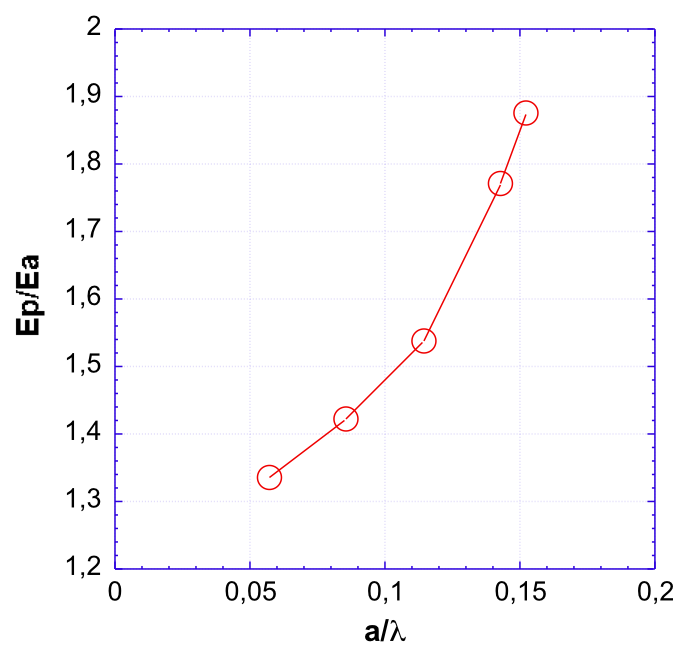}(a)}
\fbox{\includegraphics[scale=0.25]{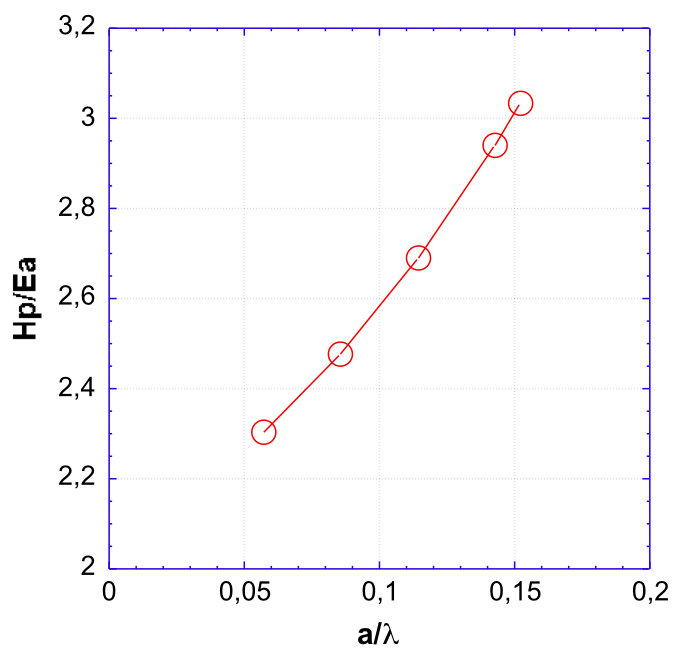}(b)}
\caption { a) $E_p/E_a$ Cavity radius as function of the $a/\lambda$ ratio. b) $H_p/E_a$ Cavity radius as function of the $a/\lambda$ ratio.}
\end{center}
 \end{figure}

 Fig. \ref{fig:Shunt} shows the main RF parameters estimations of the effective shunt impedance Rsh/m, unloaded quality factor Q and cell to cell coupling coefficient K$\%$ as function of the iris radius a/$\lambda$ obtained with the simulation code HFSS. In Fig. 3 we show the  peak electric and magnetic fields $E_p/E_a$ and $H_p/E_a$,  with $E_a$ the effective accelerating as function of the iris radius a/$\lambda$  with 5/7 semi-axes ratio of the elliptical geometry iris, for getting a satisfactory compromise with the beam dynamic requirements \cite{ref26}. As a first comment, the global behavior of all these RF parameters physically sounds since they provide a reasonable trend.

Notably, we are interested to investigate the influence of the iris on the the RF parameters in the  (0.11-0.14) a/$\lambda$  range  for getting a satisfactory compromise with the beam dynamics requirements. In this case, we observe a  variation of about 22$\%$ of the  $R_{sh}/m$,  2$\%$ of Q, 187$\%$ of K$\%$,  22$\%$ of $E_p/E_a$ and 13$\%$ of $H_p/E_a$. Since the difficult problems  limiting the linearizer performance are the ultra-high gradient operation achievable and the 36 GHz power source, we choose the geometry with a/$\lambda$ = 0.11 ratio (i.e. 1 mm iris radius) which provides a minimum ratio of $E_p/E_a$ with $H_p/E_a$ and $R_{sh}/m$ satisfactory  values if they are compared with the best estimations. This choice is also a good compromise with the beam dynamics requirements since we suppose the beam emittance is estimated to be 0.2$\mu$rad at 300 MeV \cite{ref223}.  As a result, the optimized cavity geometry of the linearizer has a rounded profile, an elliptical iris geometry of 5/7 semi-axes ratio and a iris thickness of h = 0.667 mm. The RF properties are summarized in the Table 2.

 \begin{table}[H]

\caption{RF parameters of the designed cavity}

\begin{center}

\small{ \begin{tabular}{|| c| c||}
\hline
Design parameters&  \\ 
 \hline\hline
Effective Shunt impedamce [M$\Omega$/m]& 189\\ 
\hline 
Unloaded quality factor, Q & 5778\\ 
\hline
Coupling coefficient, K$\%$&0.83 \\ 
\hline
$E_p/E_a$& 1.54 \\ 
\hline
$H_p/E_a$ [mT/MV/m]& 2.69 \\ 
\hline
 Iris thickness [mm] & 0.667 \\ 
 \hline
Cavity radius [mm]& 3.374\\ 
\hline
Cavity gap [mm]& 4.374\\ 
\hline
Iris radius [mm]& 1\\
\hline
\end{tabular}}
\end{center}
\end{table}

 \begin{figure*}[t]
 \begin{center}
 \fbox{\includegraphics[scale=0.26]{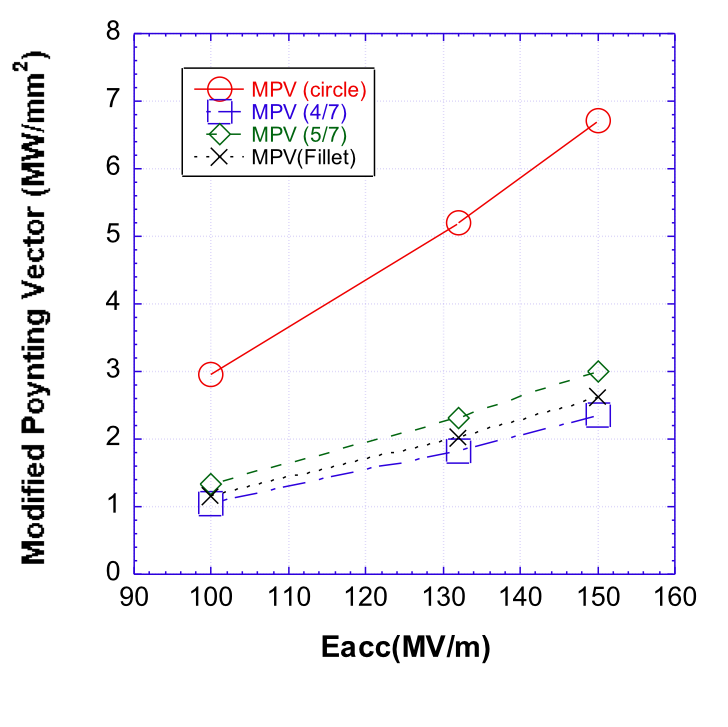}(b)}
\fbox{\includegraphics[scale=0.274]{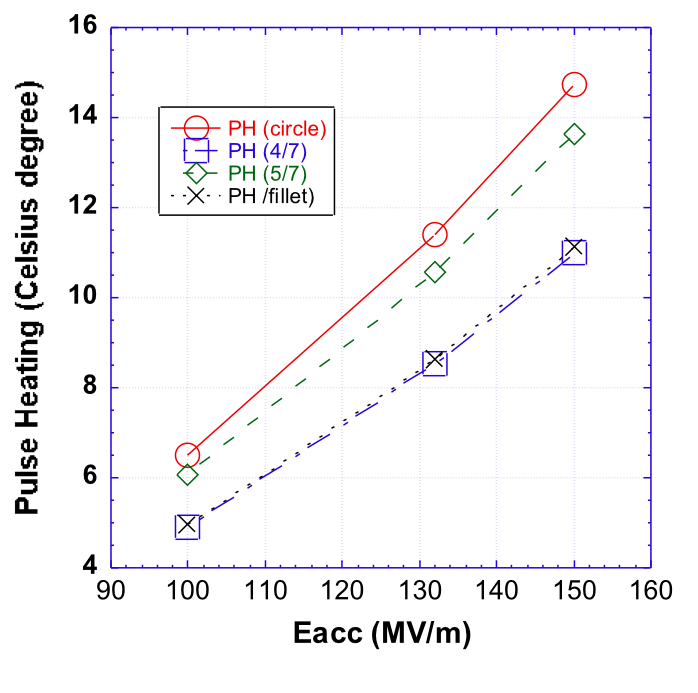}(a)}
\caption { a) Modified Poynting Vector b) Pulsed Heating as function of $E_{acc}$ }\label{fig:MPV}
\end{center}

      \end{figure*}

These RF parameters are to be chosen to achieve an ultra accelerating gradient by assuming the necessary RF power is available. However, we also note that the cavity RF parameters are conservative if compared to the usual  accelerators. 

Additional checks of the chosen cavity geometry have been carried out by estimating the quantities referring to the modified Poynting vector (MPV) and the pulse heating (PH) effects in order to estimate the allowed levels of the breakdown safety thresholds and confirm the above discussed cavity optimization. 
Figs.\ref{fig:MPV} (a) and (b) illustrate the MPV and PH  as function of the effective accelerating field by assuming a flat top 50 ns pulse length for different irises geometries, as it was already discussed for the Compact light XLS project linearizer \cite{ref26} . We observe, the MPV and the PH are in agreement with the analysis developed so far by comparing the $E_p/E_a$ and $H_p/E_a$ behavior.  At a given effective accelerating gradient, the worse case occurs for the circular iris for both  MPV and PH quantities. In all cases, we are well below the safety threshold of MPV and PH which are about  5 MW/mm$^2$ and 50 Celsius degree, respectively.  The other geometries provide  comparable values each other and they are smaller with respect to the circular iris by factor of about 2.5 for  the MPV and of the same order of magnitude  for the PH,  by confirming the above analysis. 

For the sake of completeness, the estimated MPV for the TW case is around 8 MW/mm$^2$ and thus exceeds the threshold limit of 5 MW/mm$^2$.

\section{Power source}

As we already mentioned, so far the technologies in the Ka-Band regime for accelerating structures, high power sources and modulators have been developed by getting promising results in order to reach a RF power output of (40-50) MW by using the SLED system. However, a lot of work has yet to be done in order to achieve  high power Ka-band devices with performances reproducible and reliable. Our main interest is to get at least un integrated voltage of 15 MV \cite{ref223}.  Therefore, we also are planning to finalize the linearizer design as well as engineering of the RF power source that will be able to produce up to a 10 MW input power by using a SLED system. Analytical estimations on the Gyroklystrons devices provide an  RF output of about 3 MW and with the SLED system we are able to get an RF output power of 12 MW. In addition, to create shorter RF pulses, a laser-based RF switch developed in the framework of the MIT-SLAC collaboration for the W band structures can be used to select a proper RF pulse length \cite{Kutsaev}. As a result, it is reasonable to assume an available RF power of about 11 MW in order to feed the Ka-band linearizer. Moreover, for sake of completeness, additional investigations and optimization on the 36 GHz klystron design are in progress at INFN-LNF in collaboration with Tor Vergata University of Rome \cite{refKlystron1,refKlystron2} . So far, numerical studies provided a 42$\%$ efficiency with a 20 MW RF output even if a confirmation of this estimation has to be carried out, too.

  \begin{figure*}[t]
 \begin{center}
\fbox{\includegraphics[scale=0.22]{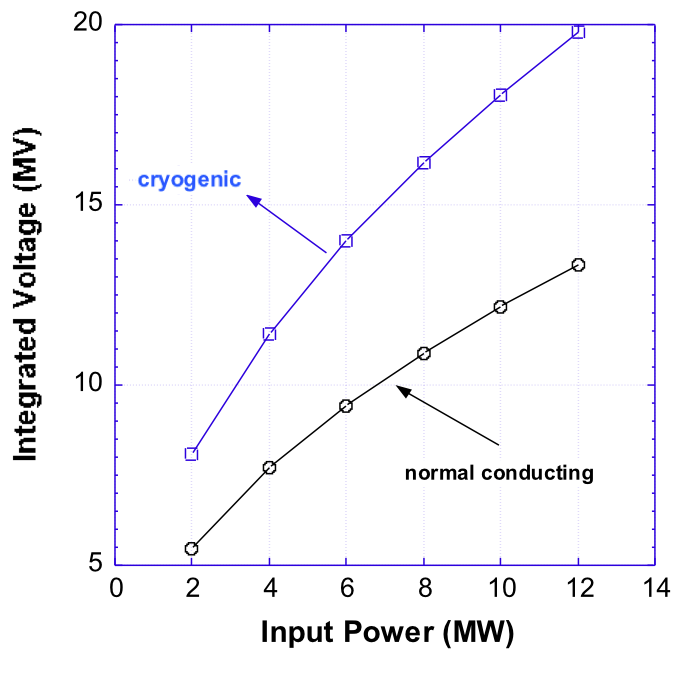}(a)}
\fbox{\includegraphics[scale=0.22]{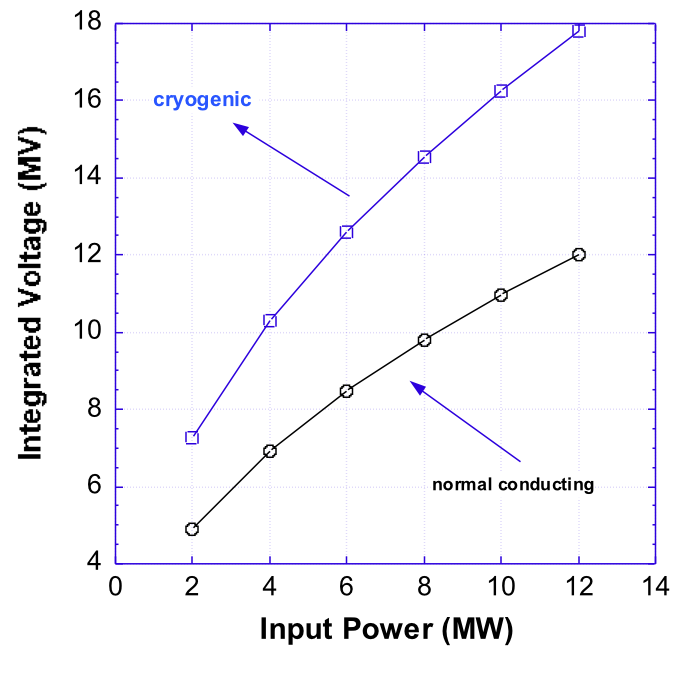}(b)}
\caption { Integrated voltage as function of the input power for  a) 1 mm and b) 1.333 mm iris radius}\label{fig:Voltage}
\end{center}

      \end{figure*}

\section{Practical achievable integrated voltage }

The linearizer has to be designed following criteria similar to those used to build SLAC X-band high gradient accelerating structures. It is our concern that this structure will operate with a surface electric field up to of 250MV/m to keep the breakdown rate probability below a reasonable value.  Due to the small dimensions of the linearizer, one of  major performance limitations are losses. 

Table 3 reports the RF parameter list of the normal conducting linearizers structures SW$_1$ and SW$_2$ with a  1 mm and 1.333 mm iris radius, respectively, for a comparison in case it is needed a larger iris for beam dynamics specific requirements.

\begin{table}[h]

\caption{RF parameters list  of the normal conducting linearizer structure}

\begin{center}

\small{ \begin{tabular}{|| c| c| c||}
\hline
Design parameters& SW$_1$&SW$_2$ \\ 
 \hline\hline
Frequency [GHz]& 34.272& 34.272\\ 
\hline 
Effective accelerating Electric field [MV/m]& 125&125\\ 
\hline 
Input power [MW]& 6.6&8\\ 
\hline
Shunt impedance [M$\Omega$/m] & 189&155\\ 
\hline
Unloaded quality factor, Q & 5778&5888\\ 
\hline
Coupling coefficient, K$\%$&0.83&2.38 \\ 
\hline
$E_p/E_a$& 1.54&1.77 \\ 
\hline
$H_p/E_a$ [mT/MV/m]& 2.69& 3.03\\ 
\hline
 Iris thickness [mm] & 0.667&0.667 \\ 
 \hline
Cavity radius [mm]& 3.794&4.878\\ 
\hline
Cavity gap [mm]& 4.374&4.374\\ 
\hline
Iris radius [mm]& 1&1.333\\
\hline
Structure length [cm]& 8&8\\
\hline
Build-up [ns]& 13.4&13.7\\
\hline
RF Pulse length flat top [ns]& 50&50\\
\hline
Repetition rate [Hz]& 100&100\\
\hline
Average RF power per meter [W/m]& 412&500\\
\hline
\end{tabular}}
\end{center}
\end{table}

By inspecting the table 3, the amount of RF parameters variation physically sounds and it is reasonable  when increasing the iris radius.
By inspecting the table 3, in order to sustain the same effective  accelerating gradient,  the SW$_2$ structure has to  be fed with a 21$\%$ power more than the SW$_1$ structure, as it is expected to be.

In normal conducting operation, we can consider an effective accelerating electric field of 125 MV/m as reasonable practical estimation, since $E_p/E_a$ and $H_p/E_a$ are enough conservative parameters by providing maximum values no higher than 222 MV/m and 379  mT respectively. On the other hand, at 125 MV/m the MPV and PH are well below the safety threshold by a factor 4 and 6  respectively. We do not foresee problems for losses per meter since the duty cycle is expected to be $5\times10^{-6}$. In both structures, the power supply can be provided by a gyro-klystron plus a SLED system. 
Fig. \ref{fig:Voltage} shows the integrated voltage as function of the input power for both 1 mm and 1.333 mm radius. At a matched input power of 8 MW, we are able to get about an integrated voltage of 12 MV and 10 MV, respectively. As a result, two separated structures are needed for providing  an integrated voltage higher than 15 MV.
Since the enhancement factor at 77 K degree is 2.2, only one structure is needed  for getting about an integrated voltage of about 15 MV.
      
\section{Conclusions}

In the framework of the UC-XFEL to be constructed at UCLA, a SW Ka-band accelerating structure has been designed for the phase-space linearization. The cavity geometry optimization design consists in a rounded profile, elliptical shape of 5/7 semi-axes ratio and with iris radius of 1 mm. The power source can be provided by a 3 MW Ka-band gyroklystron. The use of a SLED system allows to get 11 MW by assuming a compression factor of 4 in real operation to feed the linearizer. Two normal SW conducting structures provide an integrated voltage of 16 MV. If we use only one cold structure at 77~K  we are able to get an integrated voltage of about 15 MV. We also discussed the TW cavity option, although it results that for a 16 cm long structure, in order to achieve the desired integrated voltage, no RF power is available at this time due to the high-frequency attenuation $\alpha$ = 2.51~m$^{-1}$. In addition, in the case of TW option, the needed RF power for the integrated voltage, is a factor of 3 more than the SW one.
The effects of longitudinal and transverse wakes on the beam dynamics are absolutely negligible.

\end{document}